%% file: main.tex
\documentclass[sigconf,10pt]{acmart}
\makeatletter
\AtBeginDocument{\@ifundefined{@vspace@orig}{}{\let\@vspace\@vspace@orig\let\@vspacer\@vspacer@orig}}
\makeatother
\usepackage[capitalise,nameinlink]{cleveref}
\crefformat{section}{\S#2#1#3}
\crefformat{subsection}{\S#2#1#3}
\input{macros}

\title[Invariant Discovery for Networked Systems]{Invariant Discovery for Networked Systems}

\author{Hongyu Hè}
\authornote{Equal contribution.}
\affiliation{%
  \institution{Princeton University}
  \country{}
}

\author{Alexander Krentsel}
\authornotemark[1]
\affiliation{%
  \institution{UC Berkeley / Google}
  \country{}
}

\author{Sylvia Ratnasamy}
\affiliation{%
  \institution{UC Berkeley / Google}
  \country{}
}

\author{Maria Apostolaki}
\affiliation{%
  \institution{Princeton University}
  \country{}
}

\begin{document}

\copyrightyear{2026}\acmYear{2026}
\setcopyright{cc}
\setcctype{by}
\acmConference[HotNets '26]{The 25th ACM Workshop on Hot Topics in Networks}{November 16-17, 2026}{Salt Lake City, UT, USA}
\acmBooktitle{The 25th ACM Workshop on Hot Topics in Networks (HotNets '26), November 16-17, 2026, Salt Lake City, UT, USA}
\acmDOI{10.1145/3847290.3859952}
\acmISBN{979-8-4007-3029-0/2026/11}

\begin{abstract}
Invariants, the relations expected to hold among measured signals of a network, underpin applications from verification to traffic generation, telemetry imputation, and input validation.
Yet, writing them by hand demands rare expertise in both formal logic and networking.
Testing all possible combinations of telemetry variables statistically is intractable at network scale, while existing automatic tools still require a human-authored grammar of admissible invariants and learn only exact, ``hard'' rules that break under real-world noise.
Our key idea is to use an LLM to inspect variable metadata, classify counter data types, and predict candidate syntactic relationships that form a compact grammar, followed by a deterministic statistical search within that grammar.
If the initial grammar fails to recover trusted known invariants on held-out telemetry, \sys instructs the LLM to iteratively widen the grammar until recovery succeeds.
By confining the LLM to grammar induction and re-induction while delegating candidate verification to logical and statistical checks, \sys turns non-deterministic AI proposals into auditable invariants with formal guarantees within the induced grammar.
We design and prototype \sys and evaluate it on both public and production WAN telemetry, recovering expert-derived invariants with full recall and low false positives.
We close with a discussion of open problems on the path toward fully open-ended invariant discovery.

\end{abstract}




\maketitle

\input{sections/1_intro}
\input{sections/2_motivation}
\input{sections/3_approach}

\input{sections/4_eval}
\input{sections/5_agenda}

\begin{acks}
We thank Lillian Tsai and the anonymous reviewers for their insightful feedback and comments, which meaningfully improved this paper.
This work was supported by the National Science Foundation (NSF) through Grants CNS-2442625 and CNS-2319442 and Office of Naval Research (ONR) through Grant N000142612552.
\end{acks}

% \newpage
\bibliographystyle{ACM-Reference-Format}
\bibliography{reference}

\end{document}

%% file: macros.tex
\usepackage{amsfonts}
\usepackage[normalem]{ulem}
\usepackage{bbm}
\usepackage{booktabs}
\usepackage{makecell}
\usepackage{tikz}
\usetikzlibrary{arrows.meta}
\usepackage{adjustbox}
\usepackage{enumitem}
\usepackage{wrapfig}
\usepackage{subcaption}
\usepackage{xspace}
\usepackage[svgnames]{xcolor}

\usepackage{enumitem}

\usepackage{algorithm}
\usepackage[noend]{algpseudocode}
\usepackage{mathrsfs}
\usepackage{amsmath}

\usepackage{amssymb}
\usepackage[makeroom]{cancel}
\usepackage{amsthm}

\usepackage{booktabs}% http://ctan.org/pkg/booktabs

\usepackage{colortbl}

\usepackage{array}      % for p{width} columns
\usepackage{booktabs}   % for nicer rules
\usepackage{breqn}
\usepackage{tabularx}
\usepackage[framemethod=tikz]{mdframed}
\newmdenv[
    linecolor=gray,
    linewidth=1pt,
    topline=false,
    bottomline=false,
    rightline=false,
    skipabove=2pt,
    skipbelow=1pt,
    leftmargin=0,
    rightmargin=10pt,
    innerleftmargin=3pt,
    innerrightmargin=0pt,
    innertopmargin=1pt,
    innerbottommargin=2pt,
    backgroundcolor=white
]{txtframe}
\newmdenv[
    linecolor=gray,
    linewidth=1pt,
    roundcorner=4pt,
    skipabove=2pt,
    skipbelow=1pt,
    leftmargin=0,
    rightmargin=0,
    innerleftmargin=6pt,
    innerrightmargin=6pt,
    innertopmargin=5pt,
    innerbottommargin=5pt,
    backgroundcolor=gray!10
]{graytxtframe}
\newmdenv[
    linecolor=gray,
    linewidth=1pt,
    topline=true,
    bottomline=true,
    leftline=false,
    rightline=false,
    skipabove=1pt,
    skipbelow=1pt,
    leftmargin=10pt,
    rightmargin=10pt,
    innerleftmargin=2pt,
    innerrightmargin=2pt,
    innertopmargin=2pt,
    innerbottommargin=2pt,
    backgroundcolor=white
]{insightframe}

\usepackage[utf8]{inputenc}
\usepackage[T1]{fontenc}

\newif\ifshowcomment
\showcommenttrue
\newcommand{\sys}{{\textsc{Autogram}}\xspace}
\newcommand{\tocite}[1]{{\textcolor{red}{\textbf{[~]}}}}
\newcommand{\toref}[1]{\textcolor{red}{\textbf{N}}}

\newcommand{\eg}{\emph{e.g.,} }
\DeclareRobustCommand{\txtsl}[1]{%
  \ifmmode
    \text{\fontsize{9}{10}\selectfont\fontencoding{T1}\fontfamily{lmr}\fontshape{sl}\selectfont #1}%
  \else
    {\fontsize{9}{10}\selectfont\fontencoding{T1}\fontfamily{lmr}\fontshape{sl}\selectfont #1}%
  \fi}
\makeatletter
\newcommand{\etc}{etc\@ifnextchar.{}{.\@\xspace}}
\makeatother

\ifshowcomment
\newcommand{\TODO}[1]{\textcolor{red}{{[\small\textsf{{TODO: #1}}}]}}
\newcommand{\NOTE}[1]{\textcolor{orange}{{[\small\textsf{{NOTE: #1}}}]}}
\else
\newcommand{\TODO}[1]{}
\newcommand{\NOTE}[1]{}
\fi

\newcommand{\remove}[1]{}

\newcommand{\mypar}[1]{{\noindent\bf #1.\ }}

\newcommand{\circleblack}[1]{%
 \begin{tikzpicture}[baseline=(char.base)]
   \node[draw,circle,inner sep=0.5pt, fill=black, text=white] (char){\small #1};
 \end{tikzpicture}%
 }

 \usepackage{tcolorbox}
\tcbuselibrary{listings}
\usepackage{graphicx}

\newtcolorbox{promptbox}{
  colback=gray!5,
  colframe=gray!60,
  listing only,
  listing options={
    basicstyle=\ttfamily\small,
    breaklines=true,
    columns=fullflexible
  }
}

%% file: sections/1_intro.tex
\section{Introduction}
\label{sec:intro}
Network operators rely on relationships that hold between measurements in a network snapshot at any given time, known as \emph{network invariants}, for a variety of applications.
These invariants capture operational rules such as packet conservation across a router, where a node's total output equals the sum of its outgoing links.
Formal verifiers check invariants to find misconfigurations or faults~\cite{batfish2015,netkat2014,minesweeper2017,relnetverif2024,raghunathan2025layered,he2026typonet}, traffic and workload generators follow them to maintain realism~\cite{he2026netnomos,jiang2024netdiffusion,cuppers2024flowchronicle,jin2024pants,jin2027tracebleed,ustiugov2023invitro}, telemetry imputers and runtime guards use them to fill missing values and protect learning-based controllers~\cite{gong2024zoom2net,he2025lejit,gong2023towards,he2027reguard}, and data-pipeline validators check them to catch corrupt inputs before they spread~\cite{krentsel2024validating,krentsel2026crosscheck,wang2025zoomsynth}.
Making these relationships machine-checkable has caught corrupted controller inputs behind major WAN outages~\cite{krentsel2026crosscheck}, corrected drifting agricultural and IoT sensors~\cite{vasisht2017farmbeats,sharma2022lumen}, and flagged industrial process or capacity anomalies before equipment failed~\cite{feng2019invariants,kashaf2024nfty}.

Deriving usable invariants for a production network remains difficult and labor-intensive.
Writing them by hand demands expertise in both formal logic and the target system.
Even when experts know the expected invariants, real measurements carry collection errors, systematic biases, and implementation bugs that require manual data analysis to characterize~\cite{krentsel2026crosscheck}.
In production, the invariants that matter are usually \emph{soft}, meaning they hold within a tolerance across most observations.

Automatic invariant mining promises to learn these relationships directly from telemetry~\cite{he2026netnomos,hance2021swiss,yao2022duoai,yao2021distai}, yet a fundamental scaling barrier stands in the way.
A real network tracks too many variables to search through every possible combination of counters using statistical techniques alone.
To make the search tractable, existing approaches require a human expert to first author a \emph{grammar} that restricts which variable types can be grouped and compared.
Even when an expert writes that grammar by hand, existing approaches look for exact rules and miss relationships that are masked by measurement noise (\S\ref{sec:motivation}).

A grammar shrinks the search space by grouping variables into types and specifying which syntactic relationships can connect those types.
Consider a flow conservation rule $I_1$ at a router $X$, which states that total output approximately equals the sum of traffic on its outgoing links, $\text{out}_X \approx \texttt{egress}(X\!\to\!Y) + \texttt{egress}(X\!\to\!Z)$.
Expressing this rule requires four building blocks in the grammar: (1) \emph{types} so packet counts never mix with byte counts; (2) \emph{locality groups} that collect all links leaving router $X$; (3) an $n$-ary \emph{sum} whose size adapts to the router's degree; and (4) a \emph{soft comparison} that permits a bounded tolerance.
Raw counter values do not expose these building blocks, so prior systems rely on human experts to define the grammar for each dataset.

Defining a grammar requires domain intuition because raw variable names often omit structural context.
For instance, standard OpenConfig interface counters do not encode topological roles directly, so systems such as CrossCheck rely on human annotations to map counters to network structure~\cite{krentsel2026crosscheck}.
Across tabular knowledge bases and databases~\cite{kruit2020tab2know,huhtala_tane_1999}, software systems~\cite{ernst2007daikon}, and networks~\cite{he2026netnomos}, a human has always defined the space of candidate rules in advance before an algorithm searches within that space.
We observe that LLMs can now supply this domain intuition from variable names and basic metadata to classify data types and predict which types relate to one another.
Relying on an LLM, however, raises an immediate question: since LLMs are stochastic and can hallucinate, how can invariants discovered with their help remain reproducible and trustworthy?

\looseness=-1 Our key idea is to use the LLM exclusively to construct and refine the \emph{grammar}, while delegating all candidate verification to deterministic logical and statistical checks.
Asking an LLM to propose individual invariants directly produces isolated guesses with no defined hypothesis space; even if a downstream solver and statistical verifier checked each guessed rule for soundness, the search would still lack exhaustiveness and completeness, and its recall would degrade as the network grows.
Having the LLM define a grammar quantified over routers and links instead restricts the search space in a single step, which allows downstream checks to search that smaller space exhaustively across all network entities.
Because the LLM never proposes or filters individual candidate rules, three guarantees hold for whatever grammar is active (\S\ref{sec:approach}): the downstream search is \emph{deterministic and exhaustive}, evaluating every candidate in the grammar identically on every run; \emph{statistically sound}, accepting only rules that a solver proves non-trivial and that exceed a strict hold-rate threshold on the data; and \emph{monotone complete}, guaranteed to recover any expressible invariant that holds at least as often, on at least as much data, as a known invariant the operator already trusts.

We instantiate this design in \sys, where an LLM induces a bounded, typed grammar from variable names and operator metadata such as topology, role, and vendor annotations, and iteratively widens that grammar if known baseline invariants are missed.
A deterministic engine then enumerates the candidate rules in the grammar, a logical solver prunes trivial and contradictory forms, and a statistical verifier tests how consistently each surviving rule holds on telemetry.
To prevent false positives when testing thousands of candidates, \sys fits each rule's tolerance on a separate data split, demands high agreement across snapshots, and sets its minimum hold-rate threshold high enough to reject coincidental relationships on randomly shuffled data.

Running \sys on public WAN benchmarks and a large production backbone recovers every invariant previously written by experts and uncovers additional conservation and symmetry invariants that boost downstream validation (\S\ref{sec:eval}).
Two open boundaries frame our broader research agenda (\S\ref{sec:agenda}): composing discovered invariants into multi-hop relational structures~\cite{he2026typonet}, and extending verification beyond the decidability limits of linear solvers into non-linear queueing and buffer-sharing invariants~\cite{raghunathan2025layered,addanki2024reverie,zhou2026titrate}.

%% file: sections/2_motivation.tex
\input{assets/figures/grammar_space}

\section{Background and Motivation}
\label{sec:motivation}
\mypar{What a soft invariant is}
A \emph{soft invariant} is a typed relationship over named variables that holds across most snapshots of a network's data within a bounded tolerance.
Each soft invariant therefore carries a hold-rate between zero and one.
The relationships span numeric conservation invariants where a router's output equals the sum of its outgoing links, magnitude bounds where a queue length stays within its dynamic switch buffer threshold~\cite{addanki2022abm,addanki2024reverie,addanki2024credence,zhou2026titrate,raghunathan2025layered}, and logical presence pairings where a route is advertised or selected only when its next-hop interface and policy path are active~\cite{birge2022takes,birge2024tango,RouteScout}.
A soft invariant acts as an executable runtime check that returns a pass-or-fail verdict, a numeric residual, and a calibrated hold-rate on any snapshot.
Hard invariants correspond to the special case where the hold-rate equals one at zero tolerance.
Every soft invariant remains a readable symbolic formula that operators can inspect and monitors, validators, or runtime controller shields can evaluate on live telemetry~\cite{he2025lejit,he2027reguard}.

\mypar{Why soft invariants are worth discovering}
Soft invariants serve as checkable contracts that keep production systems reliable, and discovering more of them directly strengthens downstream data analysis.
In a wide-area backbone, an SDN controller computes traffic engineering allocations from an aggregated demand matrix, and validating that matrix against local router counters catches corrupted inputs that cause major outages~\cite{krentsel2026crosscheck}.
Downstream validators such as CrossCheck grow more effective as they check more independent invariants, because each additional conservation or symmetry rule tightens the constraints on noisy measurements and sharpens fault localization~\cite{krentsel2026crosscheck}.
In a multi-tenant telemetry pipeline, the conservation invariant $\text{input} = \text{output} + \text{backlog} + \text{loss}$ separates persistent packet loss from transient incast bursts caused by datacenter and distributed machine-learning workloads~\cite{he2024gcow,zhou2025incast,zhou2026starfish}.
Similarly, correlated sensor invariants catch drifting agricultural and IoT sensors before they corrupt downstream models~\cite{vasisht2017farmbeats,sharma2022lumen}, and operational invariants flag routing, capacity, or control anomalies before service degrades~\cite{feng2019invariants,birge2025bgp,kashaf2024nfty,hosseinzadeh2023cane}.
Production invariants are soft for two distinct physical reasons.
First, zero-mean measurement noise scatters residuals symmetrically around zero, which reflects aleatoric uncertainty~\cite{derkiureghian2009aleatory, hullermeier2021aleatoric, kendall2017uncertainties}.
Second, systematic bias shifts a conservation invariant by a few percent at every router, for example when interface counters include packet headers that demand matrices omit.
Both behaviors provide operational value: a near-exact rule acts as a tight correctness check, while a consistent offset diagnoses an accounting discrepancy across platforms.

\mypar{The grammar is the real bottleneck}
Defining the \emph{grammar} of candidate rules is the central bottleneck in invariant discovery because the grammar determines which relationships can be expressed and keeps the search space small enough to check (Fig.~\ref{fig:grammar-space}).
A real network exposes too many variables to test all possible combinations of counters without structural guidance.
A useful grammar shrinks that search space by identifying what each variable measures, grouping related variables into \emph{locality groups} such as all links leaving a router, allowing $n$-ary sums over those groups, and specifying which data types can be compared.
Constructing those groupings requires domain intuition about what the variable names mean.
Raw counter values do not reveal that two egress counters belong to the same outgoing link group, and hand-crafting a new grammar for every telemetry schema and network does not scale.

\mypar{Where prior methods fall short}
Existing tools require a human to define the search space by hand and struggle when measurements contain noise.
Network verifiers check whether a network satisfies a \emph{given} property~\cite{veriflow2012, yuan2020netsmc, minesweeper2017, batfish2015, relnetverif2024}, so they rely on someone else to supply the invariant first.
Classical rule-learning approaches fix the candidate templates in advance: database dependency approaches search predetermined relational forms~\cite{huhtala_tane_1999, papenbrock_hybrid_2016, fan_discovering_2011, beldiceanu_model_2012_modelseeker}, software tools such as Daikon test a static library of patterns~\cite{ernst2007daikon}, and protocol inference approaches infer boolean predicates over discrete state against an exact oracle~\cite{yao2021distai, yao2022duoai}.
The closest network invariant approach, NetNomos~\cite{he2026netnomos}, requires an expert-written grammar for each counter set and accepts only rules that hold on $100\%$ of snapshots without supporting group sums or soft tolerances.
Other statistical paradigms solve different problems: symbolic regression fits a function for a single target variable~\cite{cranmer2020discovering}, association-rule mining links discrete items~\cite{agrawal1994levelwise, stupan2022niaarm}, and Markov logic learns weights over fixed predicate schemas without per-rule tolerances~\cite{richardson2006markov}.
Recent LLM-guided search systems such as FunSearch and AlphaEvolve ask an LLM to edit candidate programs inside an evolutionary loop scored by a known objective function~\cite{romeraparedes2024funsearch, novikov2025alphaevolve, openevolve}.
Those systems query the LLM for every candidate and emit opaque code, whereas invariant discovery lacks a ground-truth scoring function and needs readable formulas that a solver can verify.
Having the LLM construct the grammar defines the entire space of readable candidates in a handful of calls while keeping the LLM out of candidate verification.

% Declare the overview early so the two-column figure can appear on page 4.
\input{assets/figures/overview}

\mypar{Why symbolic invariants complement predictive models}
Discovering explicit symbolic invariants serves a different purpose than training a black-box model to predict telemetry values.
An invariant paired with a hold-rate forms an auditable contract that operators can read and reuse across verification~\cite{raghunathan2025layered,he2026typonet}, input validation~\cite{krentsel2026crosscheck}, runtime safety enforcement~\cite{he2025lejit,he2027reguard}, and synthetic trace generation~\cite{he2026netnomos,jin2024pants}.
A black-box predictor absorbs measurement noise and systematic offsets into learned weights, which hides accounting bugs inside the model.
A symbolic soft invariant surfaces both the relationship and its residual offset, turning a two-percent gap between demand matrices and router counters into an explicit diagnostic.
Achieving that clarity requires controlling false positives without ground-truth labels while isolating the LLM's stochastic behavior so that downstream guarantees hold (\S\ref{sec:approach}).

%% file: assets/figures/grammar_space.tex
\definecolor{gorange}{RGB}{204,102,0}
\definecolor{ginv}{RGB}{31,119,180}
\begin{figure}[t]
\centering
\resizebox{\columnwidth}{!}{%
\begin{tikzpicture}[font=\footnotesize]
  % ---- universe: all relations a solver can decide (the ceiling) ----
  \draw[rounded corners=12pt, fill=blue!3, draw=gray!55, thick] (0,0) rectangle (9.0,5.9);
  \node[gray!45!black, anchor=north] at (4.5,5.78)
      {all relations a solver can decide\, (decidability ceiling)};

  % ---- widened grammar after re-induction (dashed) ----
  \draw[dashed, draw=gorange!75, line width=0.9pt, fill=gorange!7]
      plot[smooth cycle, tension=0.85] coordinates
      {(0.9,2.5)(1.7,4.2)(3.4,4.5)(5.2,3.8)(5.4,1.9)(4.1,0.8)(2.0,0.9)(0.9,1.7)};

  % ---- proposed grammar (solid) ----
  \draw[draw=gorange, line width=1.3pt, fill=gorange!17]
      plot[smooth cycle, tension=0.85] coordinates
      {(1.6,2.6)(2.1,3.8)(3.2,4.0)(4.1,3.2)(4.2,2.1)(3.3,1.5)(2.0,1.7)};

  % ---- candidate relations (hollow) ----
  \foreach \p in {(2.55,2.35),(3.55,3.15),(1.35,2.9),(4.55,2.55),
                  (6.2,4.7),(7.7,4.3),(8.3,3.2),(6.6,2.9),(5.9,3.7),
                  (7.2,5.05),(2.5,5.0),(0.55,3.5),(5.0,0.6),(8.4,4.8)}
     \draw[gray!55, line width=0.7pt] \p circle (2.4pt);

  % ---- invariants that hold (filled) ----
  \foreach \p in {(2.9,3.2),(3.6,2.15),(4.65,1.6)}
     \fill[ginv] \p circle (3.1pt);
  \foreach \p in {(7.0,4.55),(1.15,4.7)}
     \fill[ginv] \p circle (3.1pt);

  % ---- labels ----
  \node[gorange!55!black, anchor=west] at (1.95,3.62) {\bfseries Grammar $\Gamma$};
  \node[gorange!70!black, anchor=west] at (1.8,4.3) {widen via re-induction};
  \node[ginv!60!black, anchor=south, align=center] at (7.0,4.6) {beyond $\Gamma$};

  % ---- thumbnail inset of the concrete grammar ----
  \node[anchor=south east, inner sep=1.2pt, fill=white, draw=gorange, line width=0.8pt] (thumb) at (8.9,0.18)
      {\includegraphics[width=3cm]{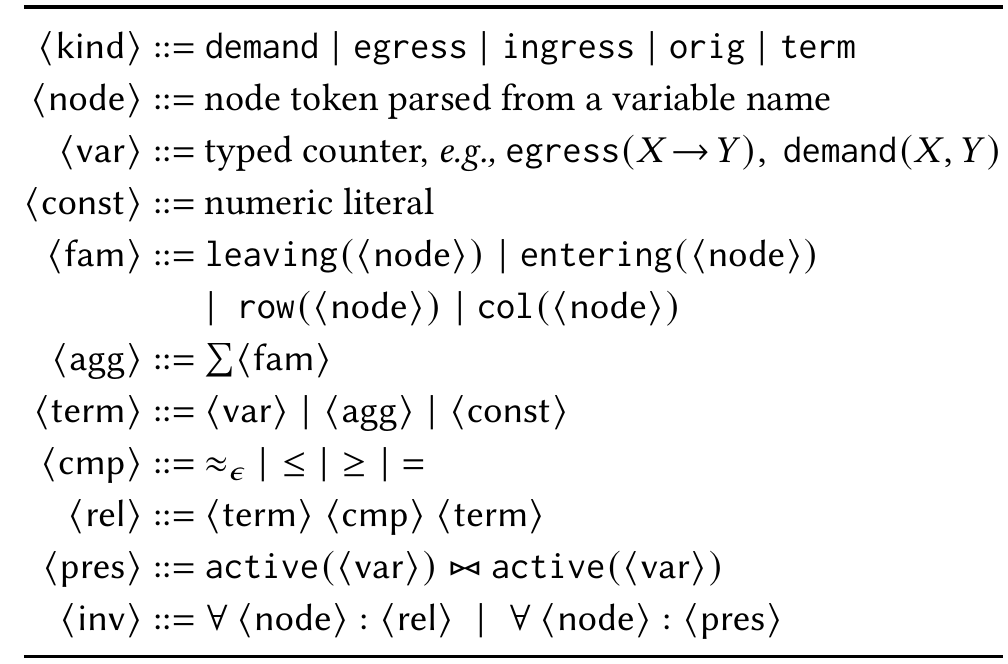}};
  \draw[gorange, line width=0.8pt, densely dotted] (3.85,1.95) to[bend left=10] (thumb.north west);
\end{tikzpicture}%
}
\caption{The grammar defines a \emph{space}.
An LLM proposes a bounded region $\Gamma$ inside the universe of all relations a solver can decide, and \sys searches only \emph{within} $\Gamma$, so a grammar \emph{confines} the hypothesis space rather than enlarging it; the inset previews the concrete grammar detailed in Table~\ref{tab:grammar}.
The data decides which points hold (\textcolor{ginv}{$\bullet$}~invariants, $\circ$~candidates), and when known invariants fall outside $\Gamma$, re-induction widens it (dashed) but never past the decidability ceiling.}
\label{fig:grammar-space}
\end{figure}

%% file: assets/figures/overview.tex
\begin{figure*}[th]
\centering
\includegraphics[width=\linewidth]{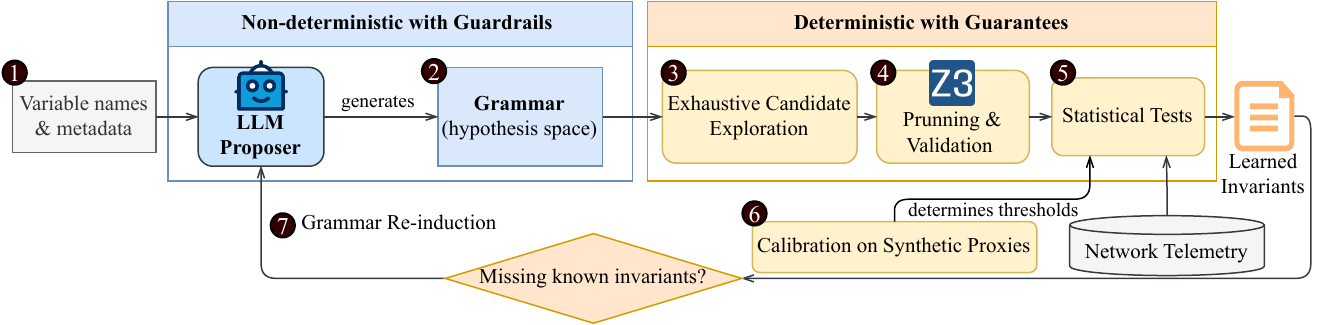}
\caption{\sys separates invariant discovery into an LLM stage that proposes a \emph{grammar} of candidate relationship templates from variable names and metadata~(\protect\circleblack{1},~\protect\circleblack{2}), and a deterministic verification pipeline driven by formal logic and statistics.
Downstream, \sys enumerates all candidates in the grammar~(\protect\circleblack{3}), prunes trivial or redundant rules with a solver~(\protect\circleblack{4}), and accepts soft invariants that exceed a data-calibrated hold-rate threshold~(\protect\circleblack{5}).
Generic thresholds are calibrated on synthetic proxies and shuffled null controls~(\protect\circleblack{6}).
If recall on a held-out split of known invariants stalls, \sys asks the LLM for a wider grammar~(\protect\circleblack{7}).
Raw telemetry reaches only the evaluator.}
\label{fig:overview}
\end{figure*}

%% file: sections/3_approach.tex
\section{Network Invariant Discovery}
\label{sec:approach}

\sys separates the single step that requires domain intuition from the downstream computation that requires formal and statistical rigor, as illustrated in Fig.~\ref{fig:overview}.

\mypar{Problem formulation}
\sys takes a dataset of network counters and measurements observed across many snapshots, paired with a small set of baseline invariants that the network operator already trusts.
The system outputs a portfolio of verified soft invariants over those variables, where each invariant carries an empirical hold-rate and a tolerance band.
Combining grammar induction with per-candidate tolerance fitting and synthetic proxy calibration allows \sys to discover soft invariants directly from noisy telemetry, where prior approaches retain only rules that hold on every snapshot~\cite{he2026netnomos,hance2021swiss,yao2021distai,yao2022duoai}.

\input{assets/tables/grammar}

\mypar{Only one step needs world knowledge}
Deciding which variable types can be grouped and compared is the only step that requires domain intuition, and that decision defines the grammar~(\circleblack{2}).
The LLM reads the variable names and metadata~\circleblack{1} and outputs a typed grammar of candidate templates without ever asserting that a specific invariant holds, as shown by the induced BNF grammar $\Gamma$ in Table~\ref{tab:grammar}.
For example, given counter names for router origination traffic, link egress traffic, and demand pairs, the LLM classifies variable kinds ($\langle\mathsf{kind}\rangle$), groups variables by router $X$ into locality groups ($\langle\mathsf{group}\rangle$) such as outgoing links $\texttt{leaving}(X)$ and demand rows $\texttt{row}(X)$, and permits comparing router totals against group sums ($\langle\mathsf{agg}\rangle$).
Because the LLM only shapes the search space by translating metadata into candidate templates~\cite{he2026typonet}, a model hallucination cannot inject an unverified invariant into the final output.

\mypar{Metadata is a rich, shareable signal}
Operator metadata such as variable names, topological roles, and vendor labels provides a compact signal for constructing the grammar~\circleblack{1}~\cite{kruit2020tab2know}.
Relying on declarative metadata interfaces brings two practical advantages for production deployments~\cite{he2024peros}.
First, raw telemetry values never leave the operator's premises during LLM queries, avoiding the source-level privacy leakage that affects even synthesized network traces~\cite{jin2027tracebleed}.
Second, because the induced grammar depends on variable names rather than measurement values, the same grammar transfers across deployments that share a telemetry schema and can be re-evaluated across multiple threshold settings after a single LLM query~\circleblack{2}.

\mypar{Balancing grammar quality through induction and re-induction}
A high-quality grammar must balance two competing goals: it must stay constrained enough to keep the search space small and fast to enumerate, while remaining expressive enough to capture the true relationships that hold in the data.
Widening the grammar captures more possible invariants, yet that expansion also increases the number of candidates to check.
To manage this tension, \sys uses a single fixed prompt template across all datasets to request a compact initial grammar~\circleblack{2}, and validates that every proposed group and type in Table~\ref{tab:grammar} maps cleanly to actual dataset columns before touching the measurements.
We then evaluate grammar expressiveness empirically by checking whether enumeration recovers known baseline invariants on held-out data; if recovery stalls, \sys prompts the LLM to widen the grammar with additional aggregations or operators~\circleblack{7}.

\mypar{An invariant's tolerance is written in its residuals}
Genuine soft invariants exhibit different noise levels and systematic offsets (\S\ref{sec:motivation}), so \sys estimates each candidate's tolerance directly from its empirical residuals.
For a candidate relation $A \approx_\epsilon B$ such as $\forall X: \texttt{orig}(X) \approx_\epsilon \sum \texttt{row}(X)$ from Table~\ref{tab:grammar}, \sys computes the dimensionless relative residual on each snapshot $t$ as
\[
v_t = \frac{|A_t - B_t|}{\max(|A_t|, |B_t|, s_0)},
\]
where $s_0 > 0$ is a small scale floor, with analogous one-sided violation residuals for $\le_\epsilon$ and $\ge_\epsilon$.
When the candidate relationship holds on the network, the relative residuals $v_t$ on most snapshots stay near zero, followed by a sharp jump to a few noisy outliers.
To avoid overfitting the error cutoff $\epsilon$ to the test data~\cite{vovk2005algorithmic,papadopoulos2002inductive,lei2018distribution}, \sys splits snapshots into a $70\%$ calibration set and a $30\%$ held-out test set.
On the $70\%$ calibration set, \sys sorts the residuals $v_{(1)} \le \dots \le v_{(m)}$, finds the sharpest jump $i^\star = \arg\max_i (v_{(i+1)} - v_{(i)})$ between the tight cluster and the outliers, and sets the error cutoff to $\epsilon = \min(v_{(i^\star)}, \epsilon_{\max})$ before measuring the hold-rate on the $30\%$ test set as the fraction of snapshots where $v_t \le \epsilon$~\circleblack{5}.
Fitting $\epsilon$ from the data captures noisy invariants such as origination demand conservation at a $68.6\%$ hold-rate in Table~\ref{tab:wan-a}, while the global cap $\epsilon_{\max} = 0.05$ prevents unrelated variables from choosing a wide cutoff to manufacture agreement.

\mypar{Logical pruning and statistical validation}
\looseness=-1 After enumerating all candidates in the grammar~\circleblack{3}, \sys applies formal logic, calibrated statistics, and algebraic dominance pruning to filter the candidate pool without relying on an LLM.
First, an SMT solver~\cite{z3} screens out trivially true candidates such as $\sum \texttt{leaving}(X) \le \sum \texttt{leaving}(X)$, self-contradictory formulas, and rules entailed by a simpler accepted invariant~\circleblack{4}.
Second, the statistical verifier checks whether each surviving candidate agrees within its tolerance band across a large majority of held-out snapshots~\circleblack{5}.
Third, to suppress uninteresting algebraic variations among statistically valid candidates, a Pareto archive collapses Z3-equivalent rearrangements into a single canonical representative, discards any rule logically entailed by a shorter accepted rule, and prunes weakened derivatives such as scaled-slack inequalities $A \le c \cdot B$ with $c > 1$ or multi-term one-sided bounds whenever a tighter approximate equality $A \approx_\epsilon B$ or atomic sign bound already holds.
Using an LLM to propose or filter individual candidates would forfeit exhaustiveness and monotone completeness because the LLM could omit valid forms or non-deterministically drop true invariants before verification.
Exhaustive search over a grammar keeps false-positive risk low across $M$ candidates because the verifier requires a large hold-rate separation rather than testing a zero-effect null at $p < 0.05$, where $M$ independent tests would produce $0.05M$ false positives on average.
Given $k$ passes out of $n$ held-out observations ($\hat{p} = k/n$), \sys computes the one-sided $(1-\alpha)$ Wilson lower confidence bound $L_\alpha(k, n)$ at $\alpha = 0.05$ and requires it to clear a high hold-rate threshold $\theta \ge 0.62$ ($z_{0.05} = 1.645$):
\[
L_\alpha(k, n) = \frac{\hat{p} + \frac{z_\alpha^2}{2n} - z_\alpha\sqrt{\frac{\hat{p}(1-\hat{p})}{n} + \frac{z_\alpha^2}{4n^2}}}{1 + \frac{z_\alpha^2}{n}} \ge \theta.
\]
For unrelated noisy counters, the baseline probability $p_{\text{null}}$ of falling within a capped $5\%$ band is only a few percent ($p_{\text{null}} \ll 0.10$).
By Hoeffding's inequality and a union bound over all $M$ candidates, the probability of even one false positive is at most $M \exp(-2n(\theta - p_{\text{null}})^2)$, which stays tiny for $M \sim 10^5$ and $n \ge 1{,}000$ snapshots.
Fitting tolerances on the $70\%$ split prevents in-sample overfitting on the $30\%$ test split, and threshold calibration~\circleblack{6} runs the full grammar against synthetic and time-shuffled null controls to reject any setting that admits a false rule on noise.

\mypar{Grammar-conditioned guarantees}
Because stochasticity is confined to grammar construction, three guarantees hold relative to whatever grammar is active.
\uline{Determinism and exhaustiveness}: given a fixed grammar and dataset, \sys enumerates every candidate template identically on every run~\circleblack{3} and retains a canonical representative of each accepted equivalence class, making every run reproducible and auditable.
\uline{Statistical soundness}: every reported invariant is proven non-trivial by the solver, evaluated on out-of-sample data splits, and required to stay within the calibrated hold-rate confidence bound across all entity groups.
\uline{Monotone completeness}: the statistical verifier applies the same hold-rate threshold to every candidate and remains blind to which rules are already known to operators.
Consequently, when the acceptance threshold is calibrated so that a trusted known invariant passes, any other rule in the grammar that holds at least as often across at least as many snapshots is guaranteed to pass as well.

%% file: assets/tables/grammar.tex
\begin{table}[t]
\centering
\footnotesize
\setlength{\tabcolsep}{4pt}
\renewcommand{\arraystretch}{1.15}
\begin{tabular}{@{}r@{\;}c@{\;}l@{}}
\toprule
$\langle\mathsf{kind}\rangle$  & $::=$ & $\texttt{demand} \mid \texttt{egress} \mid \texttt{ingress} \mid \texttt{orig} \mid \texttt{term}$ \\
$\langle\mathsf{node}\rangle$  & $::=$ & node token parsed from a variable name \\
$\langle\mathsf{var}\rangle$   & $::=$ & typed counter, \eg $\texttt{egress}(X\!\to\!Y),\ \texttt{demand}(X,Y)$ \\
$\langle\mathsf{const}\rangle$ & $::=$ & numeric literal \\
$\langle\mathsf{group}\rangle$ & $::=$ & $\texttt{leaving}(\langle\mathsf{node}\rangle) \mid \texttt{entering}(\langle\mathsf{node}\rangle)$ \\
                               &       & $\mid\ \texttt{row}(\langle\mathsf{node}\rangle) \mid \texttt{col}(\langle\mathsf{node}\rangle)$ \\
$\langle\mathsf{agg}\rangle$   & $::=$ & $\textstyle\sum \langle\mathsf{group}\rangle$ \\
$\langle\mathsf{term}\rangle$  & $::=$ & $\langle\mathsf{var}\rangle \mid \langle\mathsf{agg}\rangle \mid \langle\mathsf{const}\rangle$ \\
$\langle\mathsf{cmp}\rangle$   & $::=$ & $\approx_{\epsilon}\ \mid\ \le\ \mid\ \ge\ \mid\ =$ \\
$\langle\mathsf{rel}\rangle$   & $::=$ & $\langle\mathsf{term}\rangle\ \langle\mathsf{cmp}\rangle\ \langle\mathsf{term}\rangle$ \\
$\langle\mathsf{pres}\rangle$  & $::=$ & $\texttt{active}(\langle\mathsf{var}\rangle) \bowtie \texttt{active}(\langle\mathsf{var}\rangle)$ \\
$\langle\mathsf{inv}\rangle$   & $::=$ & $\forall\,\langle\mathsf{node}\rangle : \langle\mathsf{rel}\rangle \ \mid\ \forall\,\langle\mathsf{node}\rangle : \langle\mathsf{pres}\rangle$ \\
\bottomrule
\end{tabular}
\caption{Grammar $\Gamma$ of the phrase structure for admissible invariants that \sys induces for the CrossCheck telemetry data~\cite{krentsel2026crosscheck}.
Making locality groups ($\texttt{leaving}$, $\texttt{row}$), $n$-ary aggregation ($\textstyle\sum$), soft equality ($\approx_{\epsilon}$ within a tolerance $\epsilon$ read from the data), and presence pairing ($\bowtie$) first-class productions adaptively is what lets \sys express complex invariants that existing approaches like NetNomos~\cite{he2026netnomos} cannot.}
\label{tab:grammar}
\end{table}

%% file: sections/4_eval.tex
\section{Preliminary Results}
\label{sec:eval}

% \subsection{WAN Input Validation}
% \label{subsec:wan}

An SDN controller can reject a corrupted traffic-engineering input only when it knows which relationships a valid snapshot must satisfy, and discovering more of those invariants strengthens validation across the network~\cite{krentsel2026crosscheck}.
We use \sys to discover these invariants directly from WAN telemetry, using our experiments to test how grammar induction and soft statistical validation perform on noisy operational data.

\mypar{Setup}
We evaluate \sys on two public WAN telemetry datasets, Abilene and GEANT, packaged as CrossCheck snapshots~\cite{krentsel2026crosscheck}, and on a large production backbone we call ``WAN~prod.''
In this CrossCheck case study, each snapshot includes an aggregated traffic-engineering \emph{demand matrix} alongside per-router \emph{interface counters} measured in the dataplane.
A demand matrix is specific to this cross-layer validation case study and is not required by \sys: when given only standard router interface counters, the exact same prompt template and pipeline still discover two-end link agreement, router flow conservation, counter non-negativity, and link presence symmetry.
Our evaluation ground truth is an expert-compiled catalog of invariant categories, where CrossCheck originally used only two-end agreement and flow conservation while leaving non-negativity, zero self-demand, topology presence, and demand conservation unstated.
We compare against NetNomos~\cite{he2026netnomos}, a state-of-the-art approach that we supply with a hand-written grammar requiring over an hour of expert effort.
\sys uses the same fixed starting prompt across all three networks, taking only the dataset's counter names to induce the grammar once per network.

\mypar{Insight 1: Variable names and annotations carry the grammar}
From counter names alone, \sys recovers $100\%$ of the invariant catalog on both Abilene and GEANT, whereas NetNomos recovers under $20\%$ (Fig.~\ref{fig:wan-recall}), finding only basic non-negativity and zero-self-demand rules.
NetNomos misses the core CrossCheck invariants because two-end link agreement holds exactly on too few noisy snapshots and flow conservation cannot be written in its grammar.
Even in this benchmark, the counter names reflect structural annotations prepared for CrossCheck rather than raw router MIB strings, which are often cryptic in schemas such as OpenConfig~\cite{krentsel2026crosscheck}.
Combining cryptic variable names with the topology, role, and vendor metadata that operators already track gives the LLM the context needed to classify data types automatically.

\input{assets/figures/wan_recall}

\mypar{Insight 2: Exact approaches miss soft and aggregated invariants}
Two barriers prevent an exact approach from recovering production network invariants.
First, measurement noise and systematic bias cause genuine invariants to fail an exact $100\%$ hold-rate check: two-end link agreement holds exactly on barely one-sixth of snapshots, and flow conservation carries a consistent offset at every router.
Second, an exact approach's binary addition operator cannot express sums over variable-sized locality groups, which span $11$ links per router on Abilene and $21$ on GEANT.
\sys overcomes both barriers by inducing group-sum templates and evaluating them within data-calibrated tolerances, while pointing toward richer non-linear queueing and buffer relationships for future work~\cite{raghunathan2025layered,addanki2024reverie,zhou2026titrate} (\S\ref{sec:agenda}).

\mypar{Insight 3: Soft invariants diagnose systematic accounting offsets}
\sys discovers that the traffic a router reports originating approximately equals the row-sum of the demand matrix, and terminating traffic approximately equals the column-sum, capturing a routing-free conservation invariant directly from the counter names.
Both relationships run roughly $2\%$ short at every router because local interface counters and end-to-end demand aggregates account for packet headers and sampling intervals differently.
Rather than discarding these imperfect relationships or burying the offset inside learned neural weights, \sys reports the conservation invariant together with its $2\%$ residual band as an auditable diagnostic.

\mypar{Insight 4: Quantifying over routers and links keeps exhaustive search compact and readable}
\looseness=-1 Because \sys quantifies candidate templates over routers and links ($\forall\,\text{node}$), the number of enumerated templates depends on the schema roles rather than the node count, generating thousands of entity-level templates rather than millions of individual variable pairs.
Under the sum-only grammar in Table~\ref{tab:grammar}, enumeration at~\circleblack{3} generates $6{,}332$ raw candidates on Abilene and $2{,}972$ on GEANT and \texttt{WAN prod}, of which $3{,}923$ on Abilene and $1{,}595$ on GEANT and \texttt{WAN prod} survive complexity, admissibility, and Z3 screening at~\circleblack{4}; adding \texttt{AVG}, \texttt{MIN}, and \texttt{MAX} aggregations and scaling coefficients yields $122{,}130 \to 44{,}183$ candidates on Abilene and $91{,}908 \to 25{,}925$ on GEANT.
Statistical validation and Pareto pruning at~\circleblack{5} distill those surviving candidates into $116$ accepted rules on Abilene, $101$ on GEANT, and $41$ on \texttt{WAN prod} (Table~\ref{tab:wan-a}).
On GEANT, NetNomos outputs $1{,}645$ individual rules, including $946$ vacuous disjunctions and only $27$ matches to known invariants, whereas \sys achieves full recall in a compact portfolio that directly strengthens downstream CrossCheck validation.

\input{assets/tables/wan_a}

\input{assets/figures/wan_runtime}

\mypar{Insight 5: Production telemetry validates both scalability and precision}
\looseness=-1 To evaluate real-world scalability, we run \sys on a live production WAN with $\mathcal{O}(100)$ routers and $\mathcal{O}(1000)$ links.
Using the same prompt template and counter names alone, \sys enumerates $2{,}972$ candidates at~\circleblack{3}, retains $1{,}595$ after Z3 screening at~\circleblack{4}, and completes in ten minutes, discovering $41$ invariants across all benchmark categories including demand conservation and its $2\%$ offset (Table~\ref{tab:wan-a}).
Runtime grows gently across the evaluated networks (Fig.~\ref{fig:wan-runtime}).
When our network operator inspected the $41$ rules discovered on \texttt{WAN prod}, the operator saw no spurious invariants empirically in practice: while the public benchmarks admit a few incidental inequalities from correlated traffic, every rule accepted on the production WAN captured a genuine topological or traffic invariant, and the compact $41$-rule output lets an operator verify the entire portfolio in minutes.

%% file: assets/figures/wan_recall.tex
\begin{figure}[t]
\centering
\begin{tikzpicture}[font=\footnotesize]
  \definecolor{agblue}{RGB}{31,119,180}
  \definecolor{nngray}{RGB}{175,175,175}
  % y gridlines and ticks: 0/50/100 percent -> 0/1.75/3.5 cm
  \foreach \p/\h in {0/0,50/1.75,100/3.5}{
    \draw[gray!25] (0,\h) -- (6.0,\h);
    \node[left,gray] at (-0.05,\h) {\p};
  }
  \node[rotate=90] at (-0.72,1.75) {Recall [\%]};
  % legend (above the plot to avoid the tall bars)
  \draw[black,fill=nngray] (0.9,4.0) rectangle (1.15,4.2);
  \node[right] at (1.15,4.1) {NetNomos};
  \draw[black,fill=agblue] (3.1,4.0) rectangle (3.35,4.2);
  \node[right] at (3.35,4.1) {\sys};
  % Abilene group
  \draw[black,fill=nngray] (0.6,0) rectangle (1.5,0.693);
  \draw[black,fill=agblue] (1.7,0) rectangle (2.6,3.5);
  \node[above] at (1.05,0.693) {$19.8$};
  \node[above] at (2.15,3.5) {$\mathbf{100}$};
  \node at (1.6,-0.32) {Abilene};
  % GEANT group
  \draw[black,fill=nngray] (3.7,0) rectangle (4.6,0.5285);
  \draw[black,fill=agblue] (4.8,0) rectangle (5.7,3.5);
  \node[above] at (4.15,0.5285) {$15.1$};
  \node[above] at (5.25,3.5) {$\mathbf{100}$};
  \node at (4.7,-0.32) {GEANT};
  \draw[thick] (0,0) -- (6.0,0);
\end{tikzpicture}
\caption{\sys recovers every invariant on both networks, whereas NetNomos recovers barely a fifth because it keeps only rules that hold exactly.}
\label{fig:wan-recall}
\end{figure}

%% file: assets/tables/wan_a.tex
\begin{table}[t]
\centering
\footnotesize
\setlength{\tabcolsep}{4.5pt}
\begin{tabular}{@{}lrl@{}}
\toprule
Invariant & Hold-rate & Origin \\
\midrule
Two-end counter agreement & $96.5\%$ & CrossCheck \\
Router \& network flow conservation & $92.3$--$97.5\%$ & CrossCheck \\
Non-negativity, zero self-demand & $100\%$ & \sys \\
Topology \& demand presence symmetry & $92.4$--$97.3\%$ & \sys \\
\textbf{Demand conservation (termination)} & $\mathbf{73.2\%}$ & \textbf{\sys} \\
\textbf{Demand conservation (origination)} & $\mathbf{68.6\%}$ & \textbf{\sys} \\
\bottomrule
\end{tabular}
\caption{On a production network, \sys recovers CrossCheck's invariants and discovers deeper ones it never reported (bold).}
\label{tab:wan-a}
\end{table}

%% file: assets/figures/wan_runtime.tex
\begin{figure}[t]
\centering
\resizebox{\linewidth}{!}{%
\begin{tikzpicture}[font=\footnotesize,y=1.3cm]
  \definecolor{agblue}{RGB}{31,119,180}
  % y gridlines/ticks: runtime minutes -> y_cm = 0.08*t (0..10 min)
  \foreach \t/\y in {0/0, 5/0.4, 10/0.8}{
    \draw[gray!25,dotted] (0,\y) -- (7.2,\y);
    \node[left,gray] at (-0.05,\y) {\t};
  }
  % x ticks: log10(n) from 10 to 100 -> x_cm = 6.5*(log10(n)-1)
  \draw[gray!20,dotted] (0,0) -- (0,0.95);
  \node[below,gray] at (0,-0.05) {$\mathcal{O}(10)$};
  \draw[gray!20,dotted] (6.5,0) -- (6.5,0.95);
  \node[below,gray] at (6.5,-0.05) {$\mathcal{O}(100)$};
  \node[rotate=90] at (-0.7,0.45) {Runtime [min]};
  \node[below] at (3.25,-0.45) {Number of nodes};
  % measured points: Abilene (12, 2.2 min), GEANT (22, 3.6 min), WAN prod (100, 10 min)
  \draw[agblue,very thick] (0.515,0.176) -- (2.226,0.288) -- (6.5,0.8);
  \fill[agblue] (0.515,0.176) circle (2.2pt);
  \fill[agblue] (2.226,0.288) circle (2.2pt);
  \fill[agblue] (6.5,0.8) circle (2.2pt);
  % labels for the measured points
  \node[anchor=south,inner sep=1pt] at (0.75,0.25) {Abilene};
  \node[anchor=south,inner sep=1pt] at (2.28,0.36) {GEANT};
  \node[anchor=south east,inner sep=2pt] at (6.45,0.82) {WAN~prod};
  % axes with arrows
  \draw[thick,-{Latex}] (0,0) -- (7.4,0);
  \draw[thick,-{Latex}] (0,0) -- (0,1.05);
\end{tikzpicture}%
}
\caption{\sys discovers a network's invariants in minutes across the evaluated networks.}
\label{fig:wan-runtime}
\vspace{-1.5\baselineskip}
\end{figure}

%% file: sections/5_agenda.tex
\section{Research Agenda}
\label{sec:agenda}

\mypar{Discover more by annotating more}
\looseness=-1 Enriching the operator metadata that \sys reads directly expands the invariants it can discover.
Adding attributes that operators already track, such as router vendor or hardware generation, allows the grammar to express relationships that variable names alone cannot capture, such as an accounting offset that appears only on links between different router vendors.

\mypar{When the annotations are wrong}
\looseness=-1 \sys treats metadata as a starting signal, yet variable names and annotations often drift, conflict across teams, or contain human errors~\cite{frenay2014label}.
Inaccurate labels can skew the induced grammar, yet discovered invariants also help diagnose those very labeling errors: when a conservation invariant holds across every router except one interface, that outlier provides strong evidence that the interface is mislabeled~\cite{rekatsinas2017holoclean}.
Characterizing how much label noise the pipeline can tolerate before grammar quality degrades remains an open problem.

\mypar{Let the data shape the grammar}
\looseness=-1 Summary statistics from the telemetry itself can guide grammar construction before full enumeration begins.
A lightweight statistical pass that clusters variables with correlated time series can narrow which variable groups belong in the grammar, turning grammar induction into data-guided refinement~\cite{ernst2007daikon, zhang2002association}.

\mypar{Compose invariants into new ones}
\looseness=-1 Discovered invariants can serve as building blocks for multi-step deduction.
Chaining router-level invariants across adjacent hops can derive network-wide relationships that no single template expresses in isolation~\cite{he2026typonet,birge2024tango}.

\mypar{Proxies that do not presuppose the answer}
\looseness=-1 Evaluating grammar recall against a catalog of known invariants still depends on having a trusted seed set.
A broader calibration signal can score candidate grammars directly on downstream tasks such as telemetry imputation~\cite{gong2024zoom2net,gong2023towards}, traffic forecasting and control~\cite{papagiannaki2005forecasting,hosseinzadeh2023cane,addanki2024credence}, or trace and workload synthesis~\cite{he2026netnomos,jin2027tracebleed,ustiugov2023invitro,wang2025zoomsynth,yin2022netshare}, rewarding grammars that capture the joint structure of the telemetry without referencing known invariants.